\documentclass[final,1p,times]{elsarticle} 

\usepackage{graphicx}
\usepackage{amssymb} 
\usepackage{amsthm} 
\usepackage{lineno}

\journal{Nuclear Physics A} 

\begin{document}

\begin{frontmatter} 

\title{Open questions in QCD at high parton density}

\author{Cyrille Marquet}
\address{
Departamento de F\'\i sica de Part\'\i culas and IGFAE\\
Universidade de Santiago de Compostela, 15782 Santiago de Compostela, Spain\\
and\\
Physics Department, Theory Unit, CERN, 1211 Gen\`eve 23, Switzerland}


\begin{abstract} 

I present the state of our understanding of the QCD dynamics at play in the parton saturation regime
of nuclear wave functions. I explain what are the biggest open questions in the field, their
intrinsic interest, but also why is it important to answer them from the quark-gluon-plasma
physicists' perspective. Focusing on those aspects that proton-nucleus collisions cannot investigate
to a satisfactory degree, I show that future high-energy electron-ion colliders have the potential to
address these questions, providing thorough answers in most cases, and exploratory measurements
otherwise.

\end{abstract} 

\end{frontmatter} 


\section{Introduction: what we know}

The QCD description of hadrons and nuclei in terms of quarks and gluons consists of several components: depending on their transverse momentum $k_T$ and longitudinal momentum fraction $x$, the partons behave differently, reflecting the different regimes of the hadronic/nuclear wave function. At asymptotically small $x$, due to the growth of the parton densities, their QCD evolution becomes non-linear. The emergence of this non-linear regime, called parton saturation, in which QCD stays weakly coupled, is a fundamental consequence of QCD dynamics.

\begin{figure}[t]
\begin{center}
\includegraphics[width=0.42\textwidth]{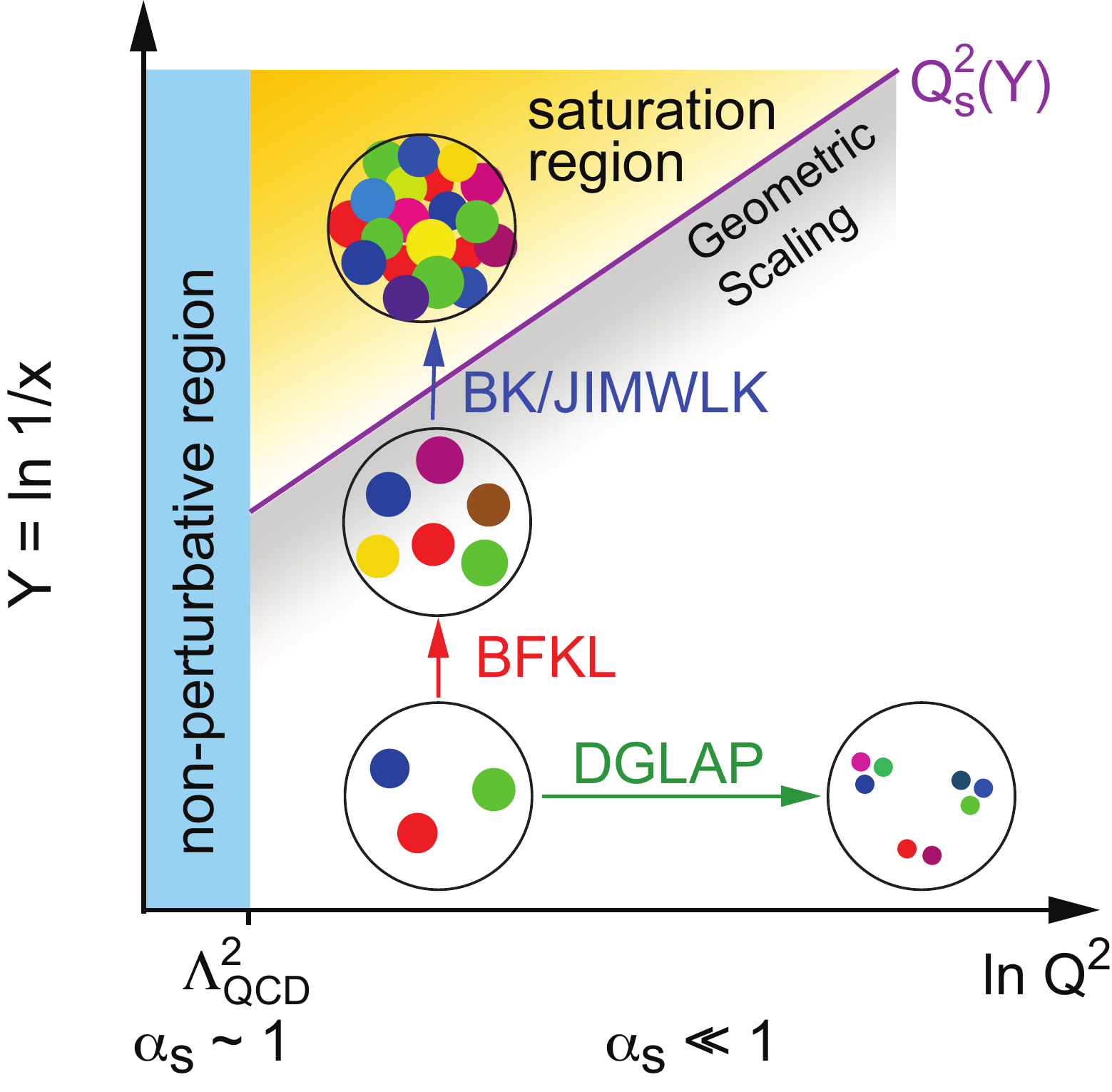}
\hfill
\includegraphics[width=0.53\textwidth]{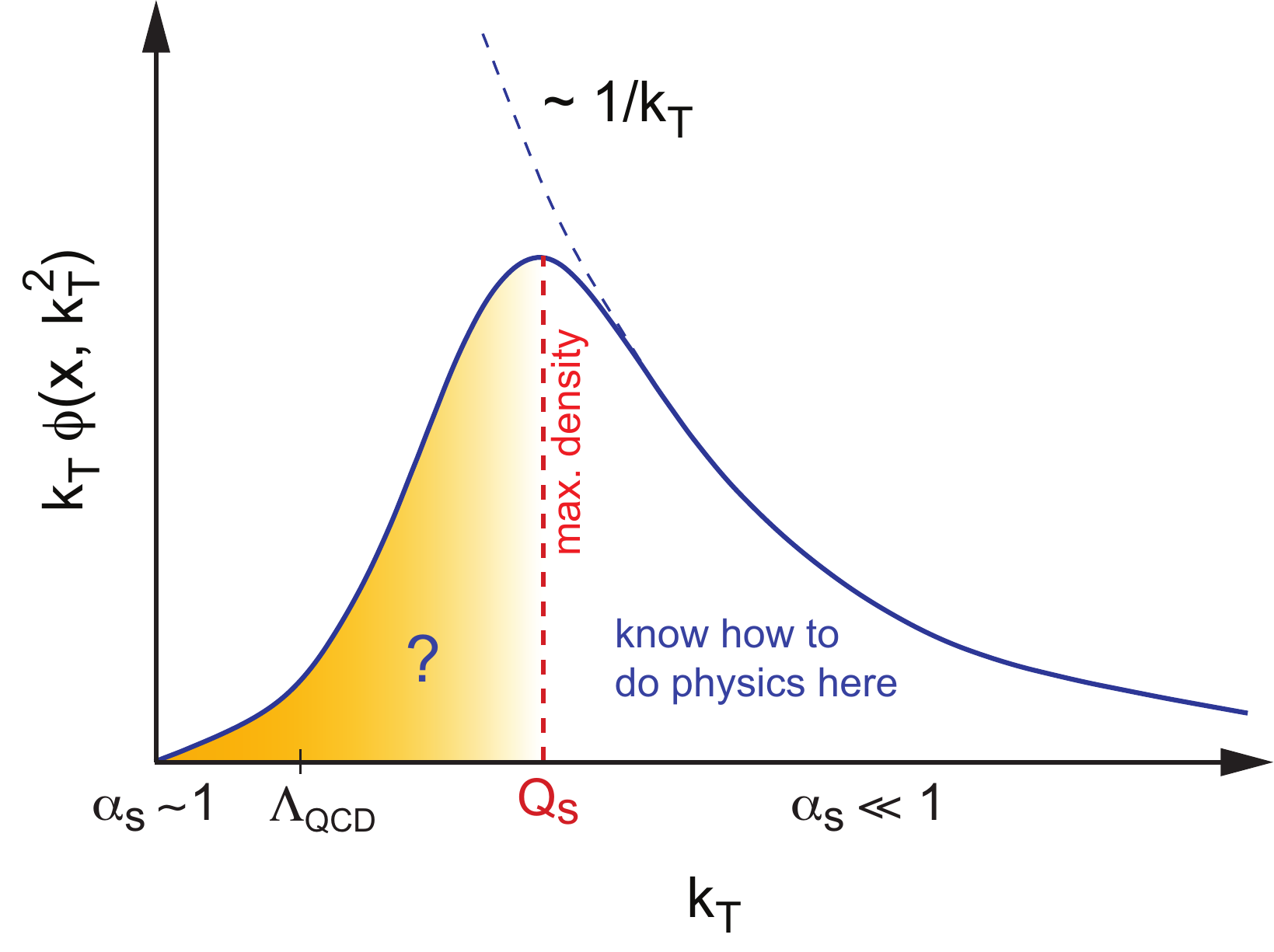}
\end{center}
\caption{Left: diagram picturing the different regimes of the hadron wave function, the saturation line separates the dilute (DGLAP) regime from the dense (saturation) regime. Right: the $k_T$ dependence of the gluon distribution at a given $x$. Most gluons carry $k_T\!\sim\!Q_s$. With decreasing $x$, $Q_s$ increases and the gluon content shifts from the unknown non-perturbative region into a regime theoretically under control ($Q_s\gg\Lambda_{QCD}$).}
\label{fig:whatweknow}
\end{figure}

The larger $k_T$ is, the smallest $x$ needs to be to reach the saturation regime. As pictured in Fig.~\ref{fig:whatweknow}, this means that the separation between the dense and dilute regimes is characterized by a momentum scale $Q_s(x)$, called the saturation scale, which increases as $x$ decreases. Non-linear effects affect not only how the parton densities evolve to smaller x, but also how the partons interact in a scattering process. Dilute partons (with $k_T\!\gg\!Q_s(x)$) scatter incoherently, as described by the leading-twist approximation of QCD. But when the parton densities are large ($k_T\!\sim\!Q_s(x)$), partons scatter collectively, and particle production becomes non-linear as well.

The Color Glass Condensate (CGC) effective theory has emerged as the best candidate to approximate QCD in the saturation regime, both in terms of practical applicability and of phenomenological success \cite{albacete}. In this framework, the energy dependence of the saturation scale, and more generally that of physical observables, can be computed from first principles, provided $Q_s\gg\Lambda_{QCD}$. This condition is better realized with higher energies (as they open up the phase space towards lower values of $x$), and with nuclear targets (since roughly $Q_s\!\sim\!A^{1/3}$).

In practice, the predictive power of the CGC depends on the level of accuracy of the calculations (leading-order vs. next-to-leading order) and on the amount of non-perturbative inputs needed (initial conditions to the small-$x$ evolution, impact parameter dependence). Electron-ion (e+A) collisions provide the best option to reduce these uncertainties and improve the CGC.

\section{Open questions: what we would like to know}

\subsection{A big open question: is parton saturation relevant at today's collider?}

In other words, can one get away with using a purely-linear gluon distribution with a necessary non-perturbative ad hoc cutoff, or do we need to properly take into account dynamics at $k_T\sim Q_s$, as suggested in Fig.~\ref{fig:whatweknow} ? On top of being much more satisfactory from the theoretical point-of-view, is the latter solution necessary ? 

My argument in favor of a positive answer is that the CGC phenomenology is successful for every collider process that involves small-x partons and $k_T\sim Q_s$, i.e. for a broad range of observables (multiplicities in p+p, d+Au, Au+Au and Pb+Pb, forward spectra and correlations in p+p and d+Au, total, diffractive and exclusive cross sections in e+p and e+A, ...). For completeness, let me say that the CGC is not yet widely accepted for two main reasons. First, because the applicability of the CGC can be questioned when values of $Q_s$ start to drop below 1 GeV (e.g. for p+p and peripheral d+Au collisions at RHIC). Second, because for each of these observables, there are alternative explanations (most of these alternatives are legitimate, but I should warn that there exist descriptions referred to as alternatives that are merely saturation models in disguise).

As we shall demonstrate below, electron-ion colliders (EICs) can provide smoking-gun measurements to answer the question, something that likely cannot be done with proton-nucleus (p+A) collisions, let alone nucleus-nucleus (A+A) collisions.

\subsection{Bigger open questions}

\noindent {\it How fast is the transition from the saturation regime to the high-$p_T$ (leading-twist) regime ?}

The most up-to-date non-linear QCD evolution equations, such as the running-coupling Balitsky-Kovchegov (rcBK) equation, do not contain the DGLAP limit, hence after some evolution down in $x$ (reflected in the data at forward rapidities), $R_{pA}$ predictions reach unity only at unrealistically large values of $p_T$. This is an open question, to which p+A collisions at the LHC should have already provided some answers, once electron-nucleus (e+A) data become available.

\noindent {\it What is the impact parameter dependence of the nuclear gluon density and saturation scale?}

This has always been the main non-perturbative input in CGC calculations. In the case of a proton, using an impact-parameter averaged saturation scale is enough most of the time, but in the case of a nucleus it is not. What is done in the most advanced CGC phenomenological studies, is to treat the nucleus as a collection of Woods-Saxon distributed CGCs, and to evolve (down in x) the resulting gluon density at different impact parameters independently. But is this good enough ? Even though in order to describe p+A collision data it seems it is, conceptually it is not good enough, and e+A collisions are crucial to answer this question by precisely imaging the transverse structure of nuclei at small-$x$.

\subsection{The biggest open questions}

There are more fundamental questions, still unexplored, in the field of QCD at high parton density. How does the transition from the saturation regime to confinement happen? Does the QCD coupling run with $Q_s$? Are classical fields still the right degrees of freedom? What are the universality properties of the saturation regime? p+A and e+A collisions offer special opportunities to explore this many-body system of strongly-correlated gluons. In the following, I focus on what is unique to e+A collisions. Great opportunities exist as well with p+A collisions, presented for instance in \cite{Salgado:2011wc,armesto-dumitru}.

\section{Why QGP physicists should care}

\begin{figure}[t]
\begin{center}
\includegraphics[width=0.34\textwidth]{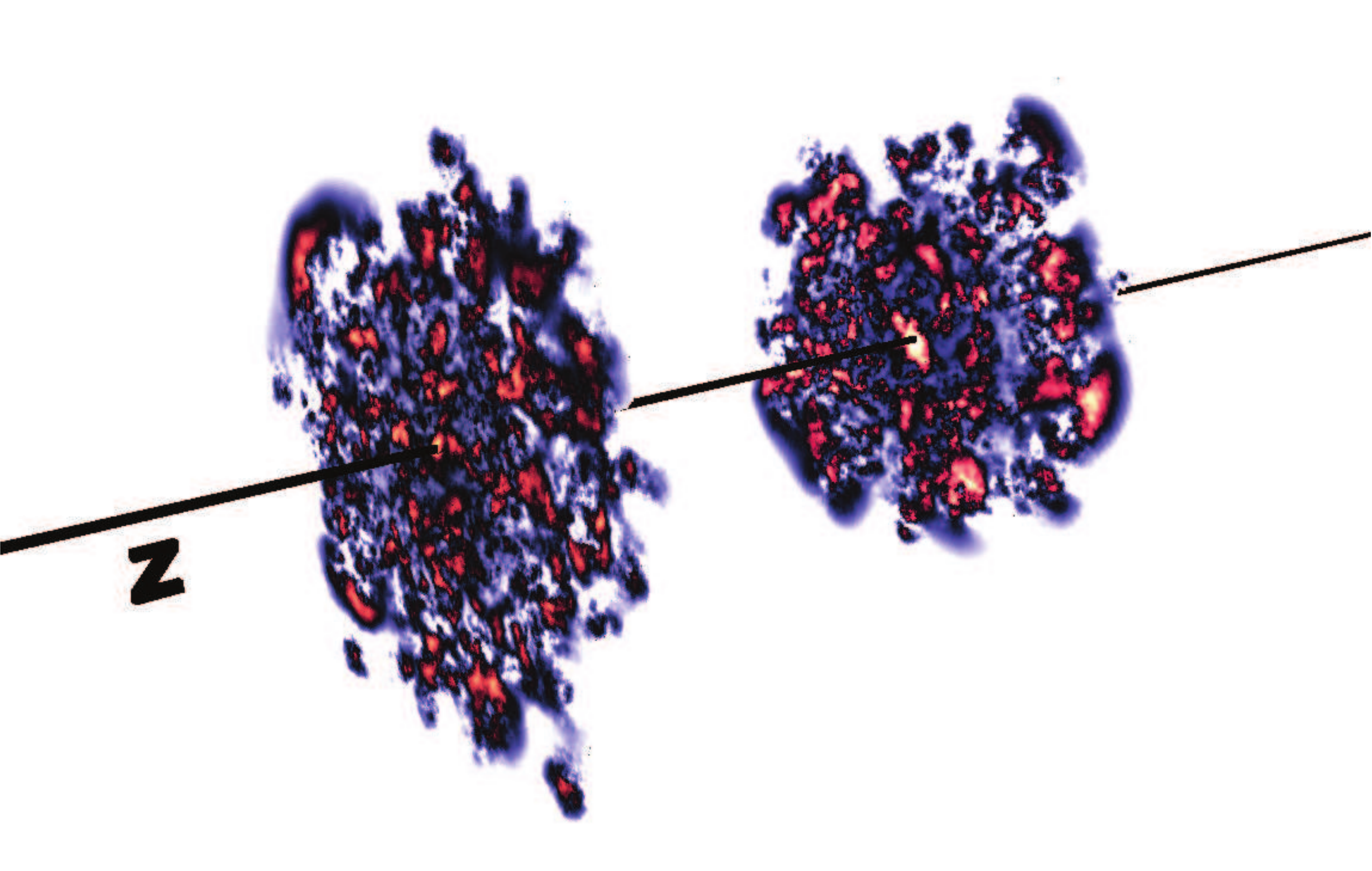}
\hfill
\includegraphics[width=0.3\textwidth]{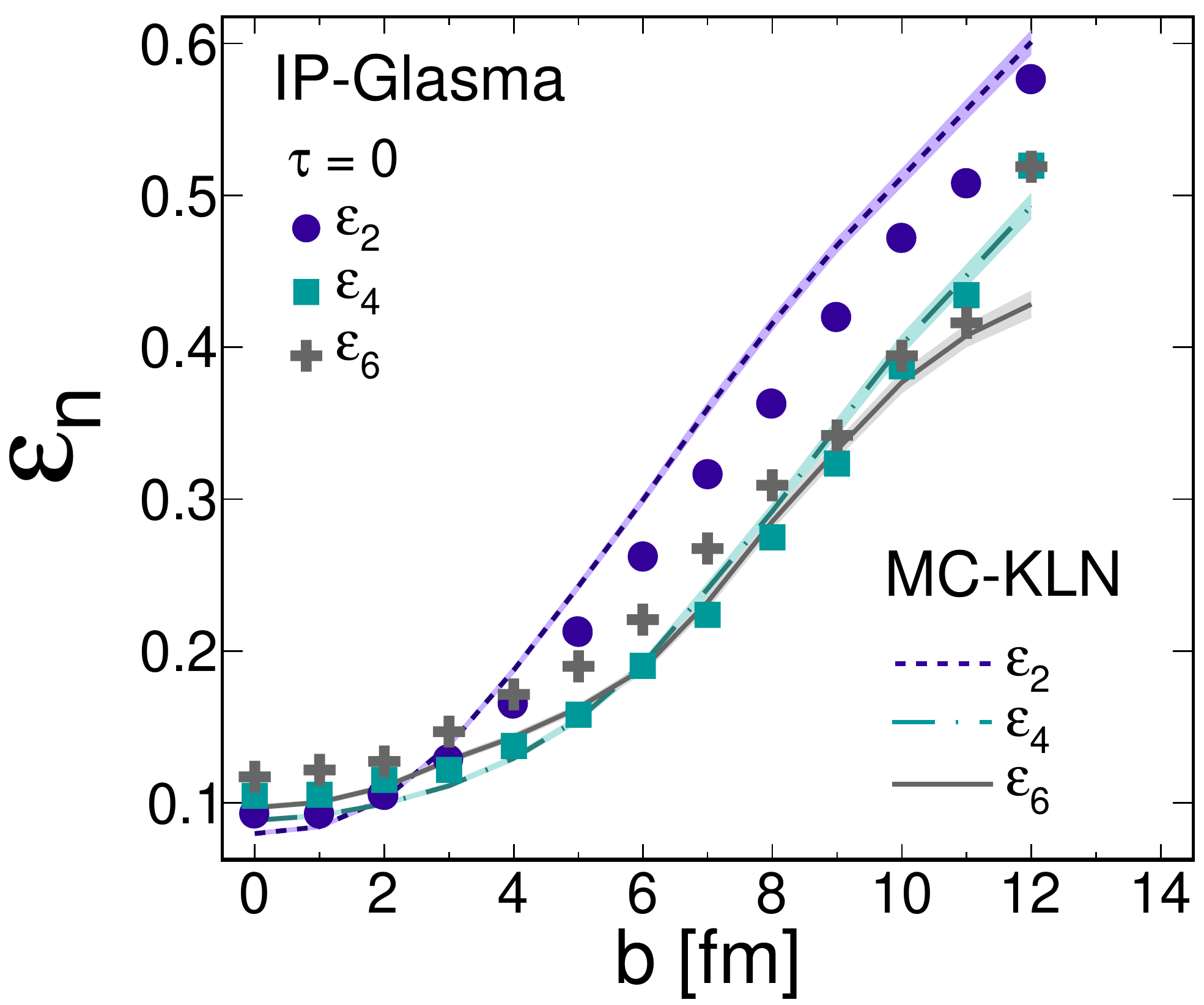}
\hfill
\includegraphics[width=0.31\textwidth]{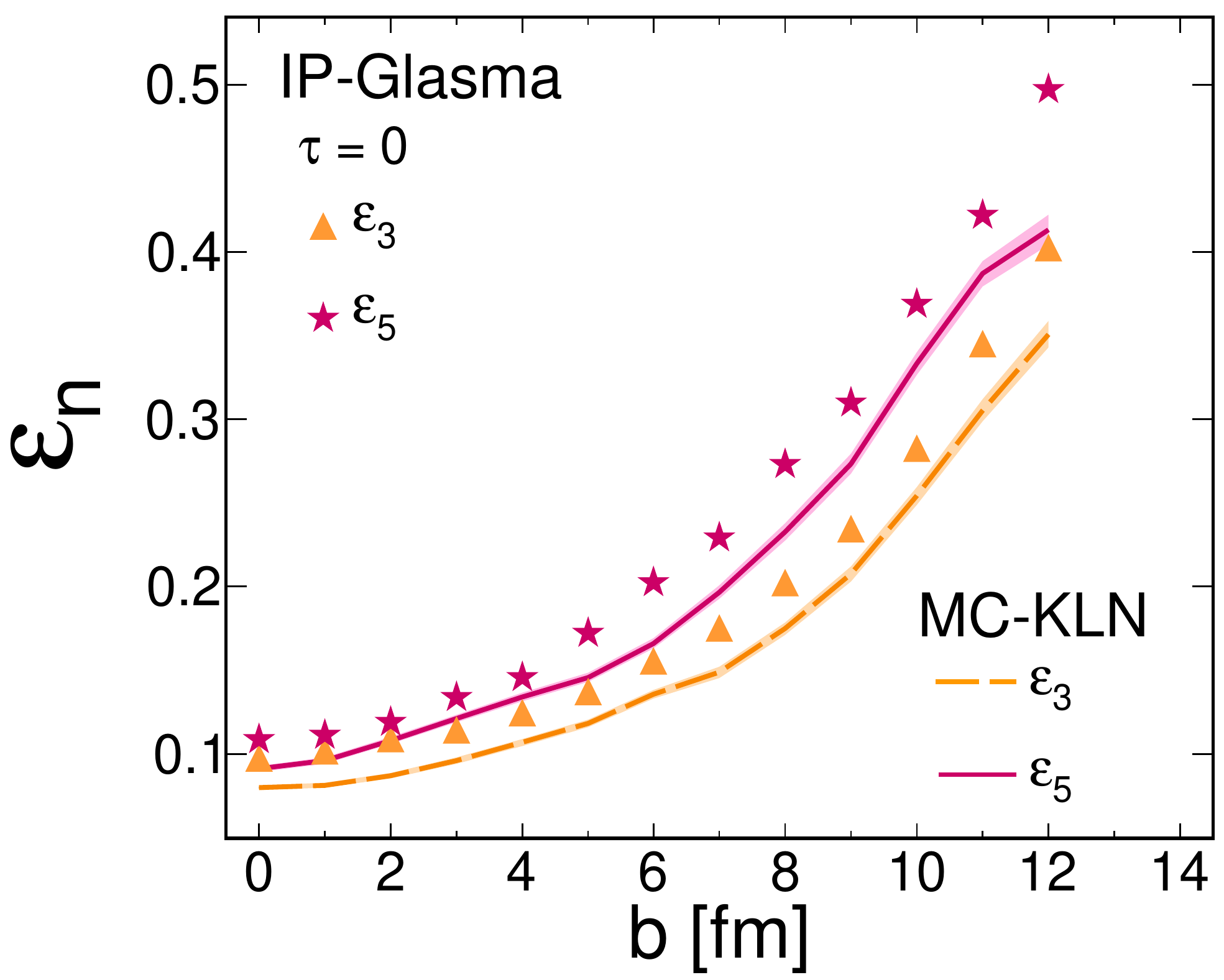}
\end{center}
\caption{Left: picture of two colliding nuclei in heavy-ion collisions, and of their fluctuating small-x gluon fields. Right: from \cite{Schenke:2012hg}, predictions of two CGC-motivated models for the first even and odd eccentricity harmonics, crucial inputs to QGP evolution models. Tests of these predictions in e+A collisions are needed to reduce the uncertainties.}
\label{whyyoushouldcare}
\end{figure}

It has recently become evident that bulk observables in heavy-ion collisions reflect the properties of the initial state as much as those of the final state, and in particular as those of the hydrodynamical evolution of the quark-gluon plasma (QGP) \cite{dusling-petersen-song-wiedemann}. As initial-state studies have historically been given less importance overall, we have reached a point where the main source of error in the extraction of medium parameters (e.g. $\eta/s$) is our insufficient understanding of the initial colliding nuclei, and more precisely of the fluctuations of the initial small-x gluon fields.

Actually, new sources of uncertainties keep emerging, and for instance as shown in Fig.~\ref{whyyoushouldcare}, even two CGC models predict different eccentricities \cite{moreland-schenke}. QGP properties cannot be precisely extracted from data without a proper understanding of the initial state. e+A collisions provide the ideal access to a precise picture of the nuclear wave function, and in particular of the part which controls soft particle production in heavy-ion collisions: its small-x component.

\section{e+A measurements: highlights}

\begin{figure}[t]
\begin{center}
\includegraphics[width=0.45\textwidth]{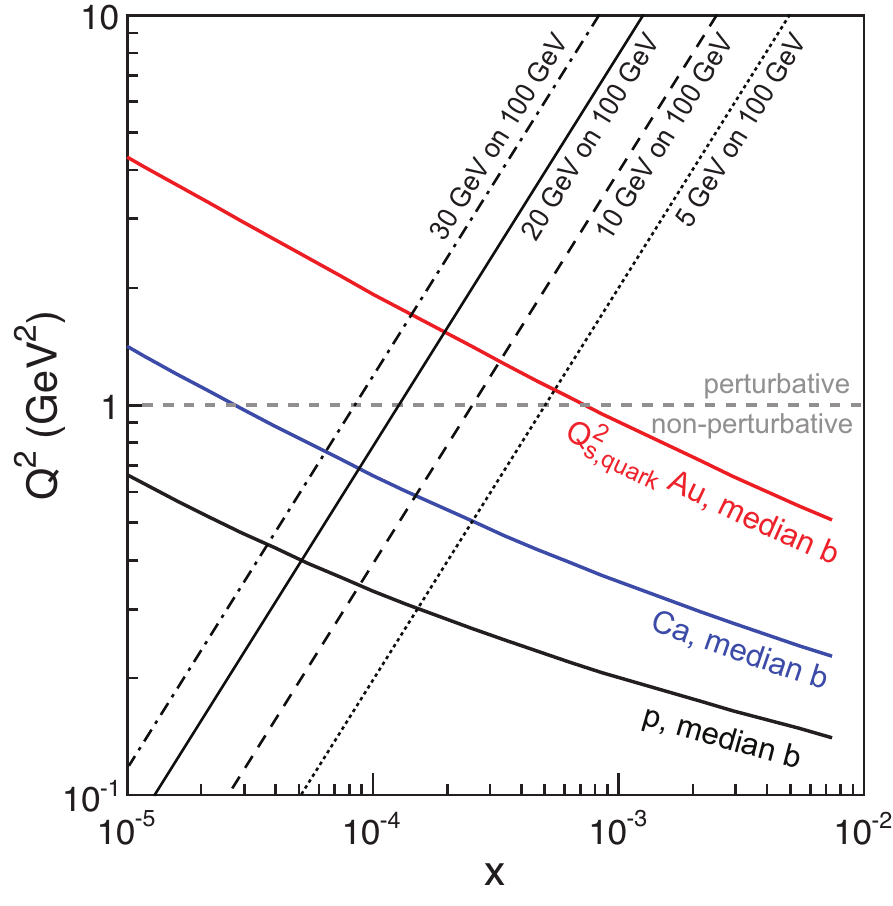}
\hfill
\includegraphics[width=0.5\textwidth]{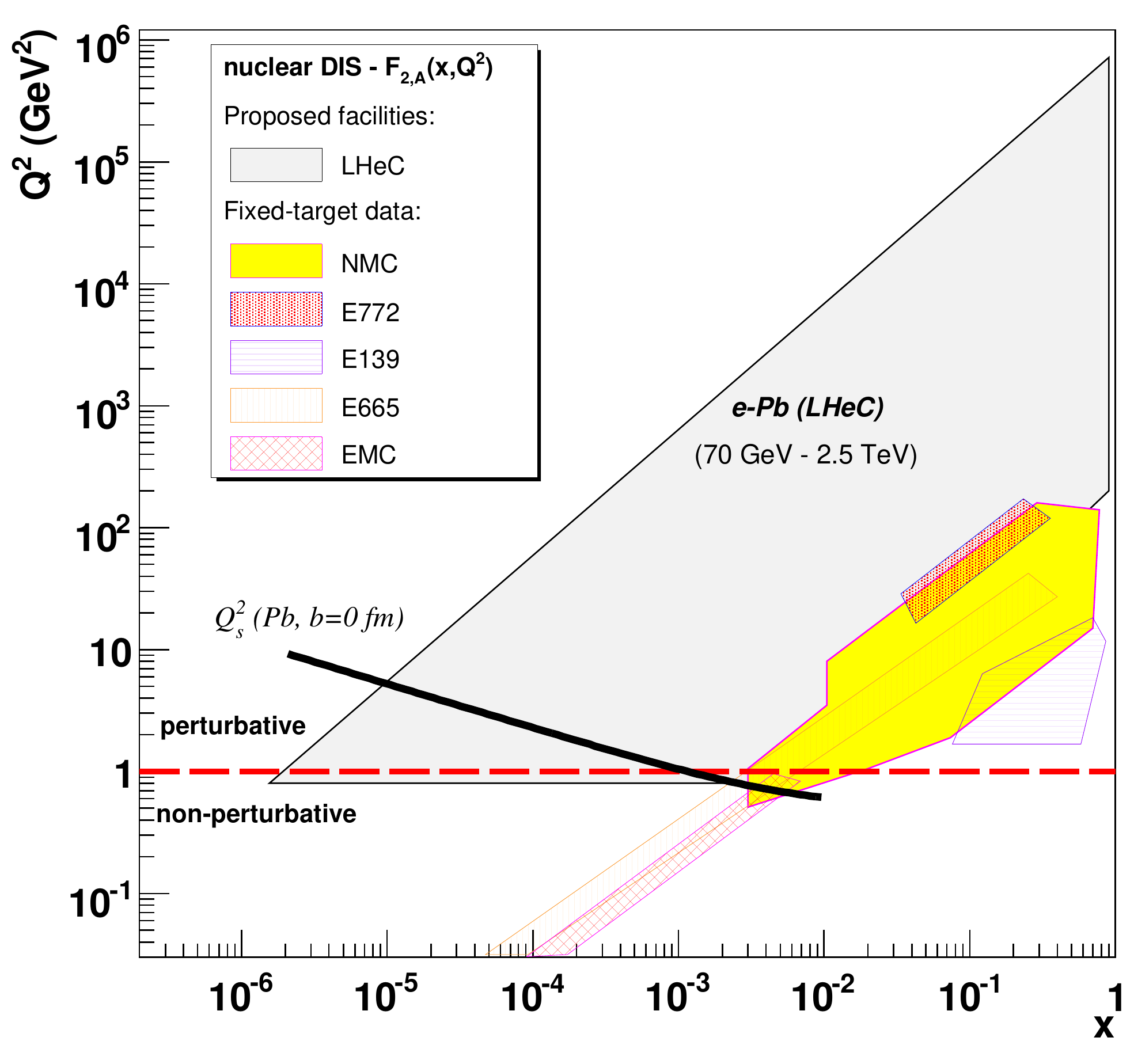}
\end{center}
\caption{Kinematic coverage of the EIC (left, from \cite{Boer:2011fh}) and LHeC (right, from \cite{AbelleiraFernandez:2012cc}) compared to different nuclear saturation scales. At the EIC, stage-2 is clearly needed to probe the saturation region of gold nuclei with $Q^2>1$ GeV$^2$.}
\label{kinematiccoverage}
\end{figure}

Fig.~\ref{kinematiccoverage} shows the kinematic coverage of the proposed Electron Ion Collider (EIC) and Large Hadron electron Collider (LHeC) in terms of Bjorken-$x$ and photon virtuality $Q^2$. These two proposals are complementary: the EIC can study the A dependence while the LHeC can reach lower $x$ values. More details about these two proposals are given in \cite{deshpande}. In the following, a selection of deep inelastic scattering (DIS) measurements that could be performed at these facilities is presented, while further examples can be found in \cite{armesto-dumitru,deshpande,lee-stasto}.

Note that not all processes require $Q^2\!\sim\!Q_s^2$ in order to be sensitive to saturation effects. While the LHeC is in general more powerful, there exist tailored observables that can provide smoking guns already at EIC energies.

\begin{figure}[h]
\begin{center}
\includegraphics[width=0.47\textwidth]{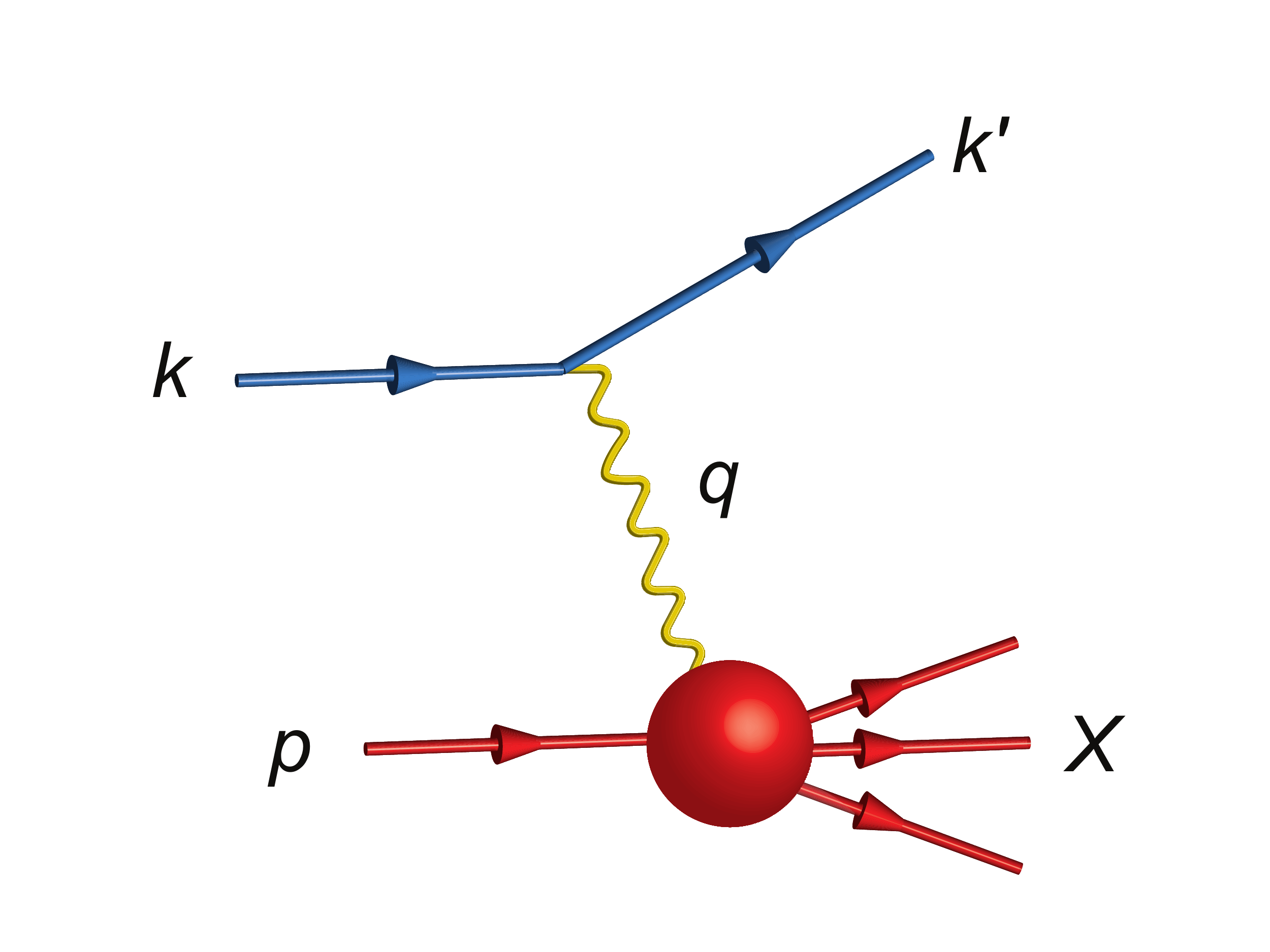}
\hfill
\includegraphics[width=0.46\textwidth]{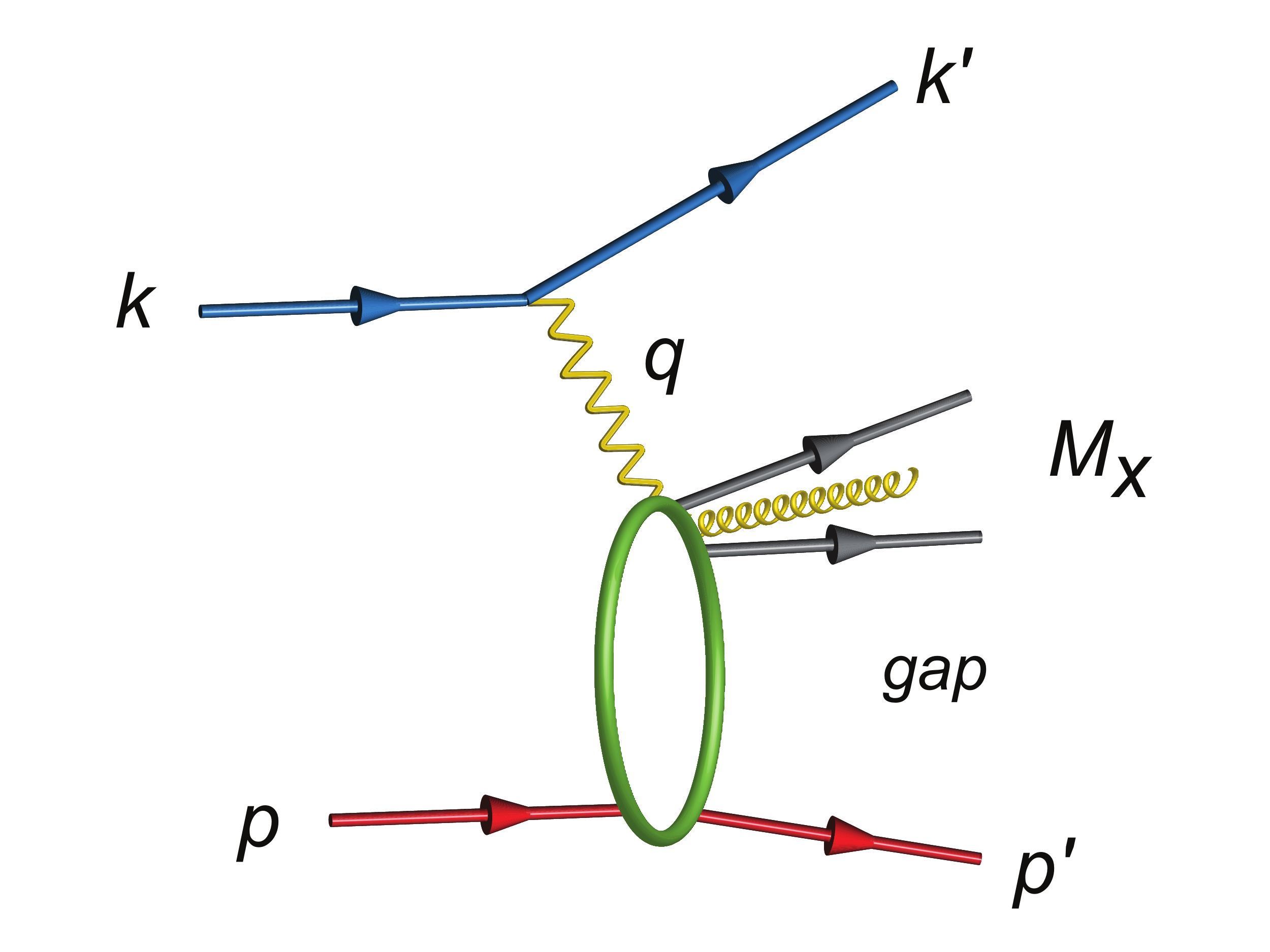}
\end{center}
\caption{Left: inclusive DIS, with photon virtuality $Q^2=-q^2$, $W^2=(p+q)^2$, and (Bjorken-)$x=Q^2/(W^2+Q^2)$. Right: diffractive DIS, characterized by two additional kinematic invariants $M_X^2=(q+p-p')^2$ and $t=(p-p')^2$.}
\label{diagrams}
\end{figure}

\subsection{Inclusive DIS}

\begin{figure}[t]
\begin{center}
\includegraphics[width=0.9\textwidth]{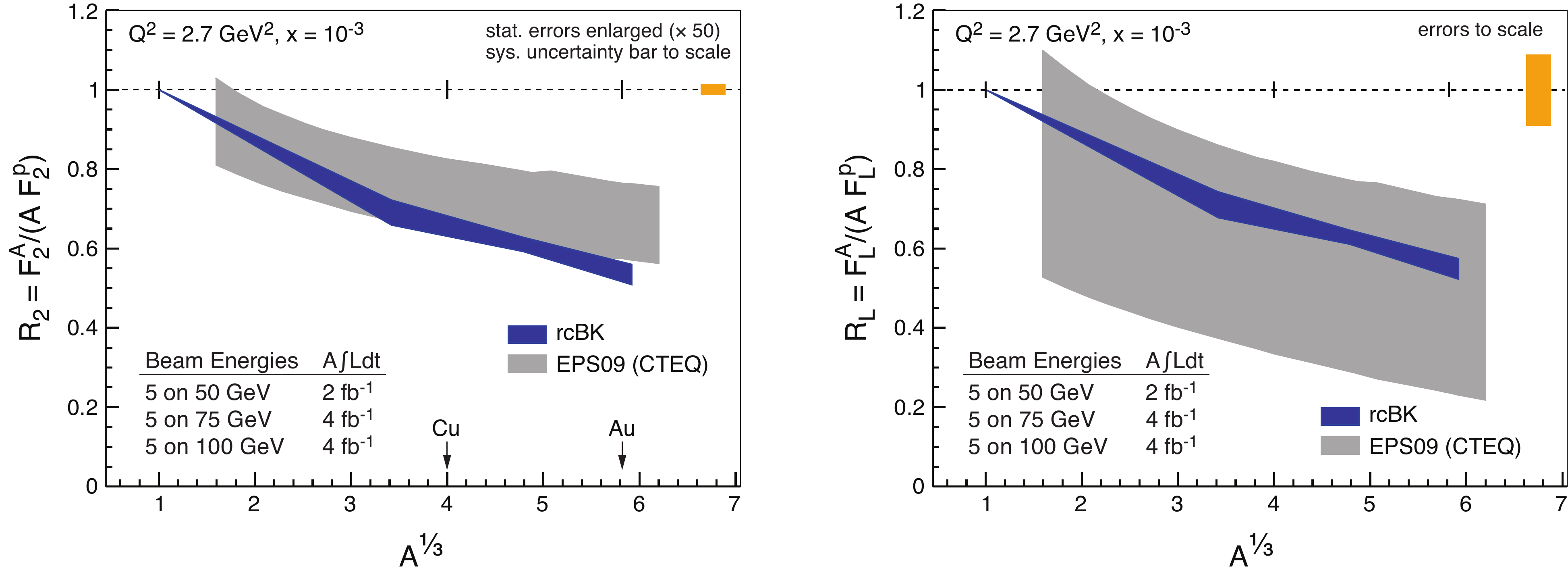}
\end{center}
\caption{EPS09 and rcBK predictions (from \cite{Deshpande:2012bu}) for $F_2^A/(AF_2^p)$ (left) and $F_L^A/(AF_L^p)$ (right) as a function of $A^{1/3}$. Expected experimental errors are also indicated on the plots for different EIC energies.}
\label{F2AandFLA}
\end{figure}

Inclusive DIS is pictured in Fig.~\ref{diagrams} (left). This process consists in measuring the total $\gamma^*\!+\!A$ cross section, a linear combination of two observables: the $F_2$ and $F_L$ structure functions. These provide access to the integrated (with respect to the transverse dynamics) distributions of quarks and gluons respectively.

Fig.~\ref{F2AandFLA} shows predictions for the ratio of nuclear to proton structure functions, as a function of $A$. The approach based on non-linear small-x evolution (rcBK \cite{Albacete:2007yr}) is predictive, and therefore would undergo a stringent test. If successful, then a further question can be asked: can the linear EPS09 approach \cite{Eskola:2009uj} (whose errors will be reduced by p+A constraints once LHC data are available) simultaneously accommodate $F_2$ and $F_L$ e+A data, if saturation sets in according to the rcBK prediction? If not, then the latter would provide the unique data description.

At the LHeC, the combined $F_2$ and $F_L$ measurements have that potential to discriminate between linear and non-linear QCD evolution. At the EIC, studies are ongoing to determine whether that is also true. In any case, it is the combination of $F_2$ and $F_L$ which makes the $\gamma^*\!+\!A$ total cross section potentially a smoking-gun measurement. Precisely measuring $F_L$ is therefore crucial, and this requires an e+A collision energy scan.

 \subsection{Diffractive DIS}

Diffractive DIS is pictured in Fig.~\ref{diagrams} (right). This measurement is identical to that of the total cross section, with the extra requirement that the target escapes the collision intact. Such processes were a surprising QCD feature at HERA: a proton in its rest frame hit by a 25 TeV electron remains intact 10\% of the time! Interestingly enough, they are naturally understood in QCD when non-linear evolution is taken into account, they are actually subject to strong non-linear effects even for $Q^2$ values significantly bigger then $Q_s^2$. For instance at HERA, the NLO DGLAP description of diffractive structure functions breaks down already below $Q^2\sim 8$ GeV$^2$.

\begin{figure}[t]
\begin{center}
\includegraphics[width=0.5\textwidth]{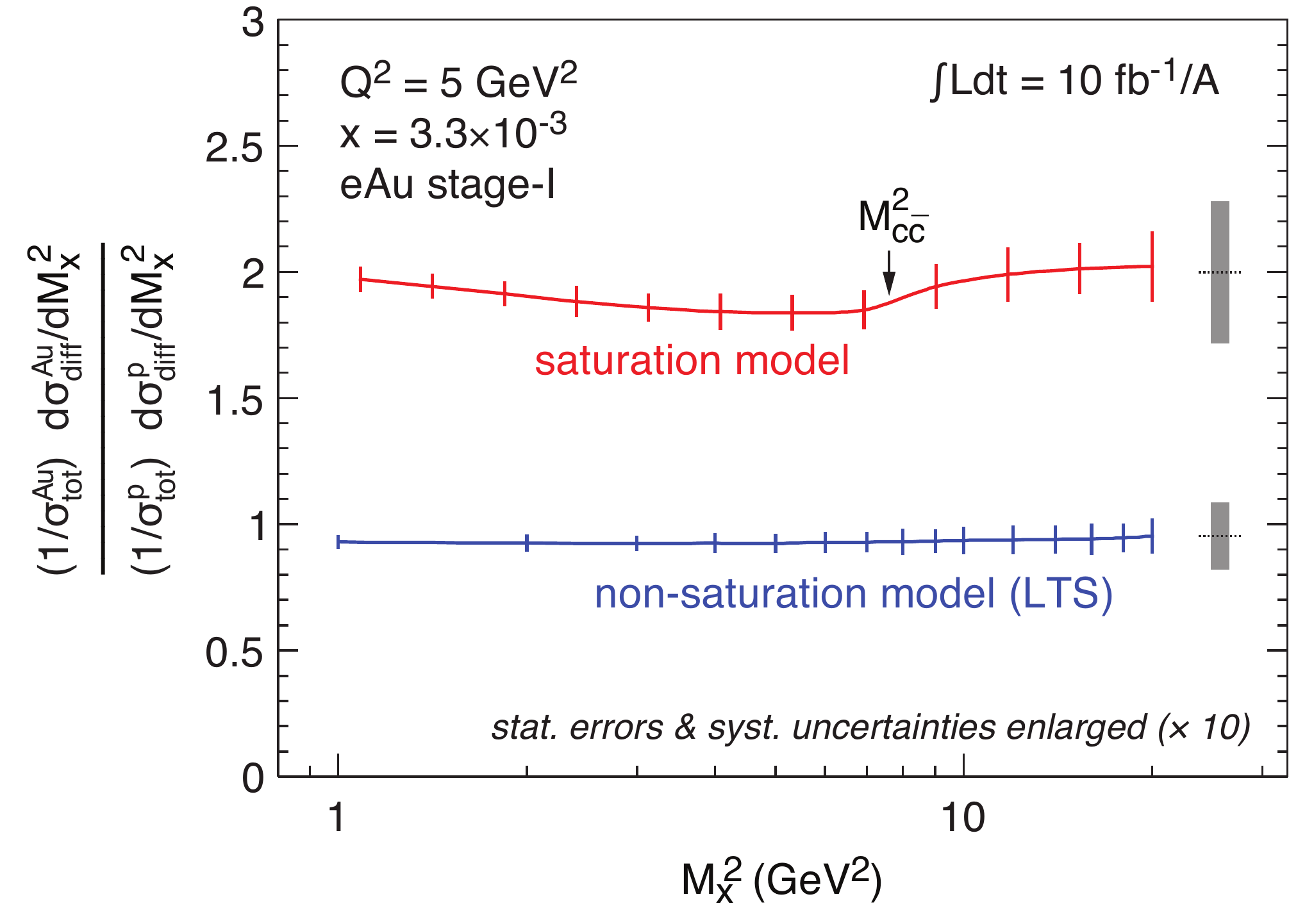}
\end{center}
\caption{Predictions (from \cite{Deshpande:2012bu}) for the gold to proton ratio of $\sigma_{diff}/\sigma_{tot}$ as a function of $M_X^2$, along with expected experimental errors at EIC energies (stage-1).}
\label{doubleratio}
\end{figure}

As shown in Fig.~\ref{doubleratio}, in e+A collisions the amount of diffractive events will be a smoking gun for parton saturation. Indeed, when non-linear QCD evolution is taken into account, one predicts that the ratio $\sigma_{diff}/\sigma_{tot}$ is a factor two larger in e+Au than in e+p collisions (in the low $M_X$ range) \cite{Kowalski:2008sa}, while such an enhancement is not seen with only linear evolution \cite{Frankfurt:2003gx}. This would be a clean and unambiguous signal of saturation already at the lowest EIC energies. Note that this enhancement is specific to e+A collisions, there is no p+A equivalent.

\subsection{Diffractive VM production and DVCS}

\begin{figure}[h]
\begin{center}
\includegraphics[width=0.45\textwidth]{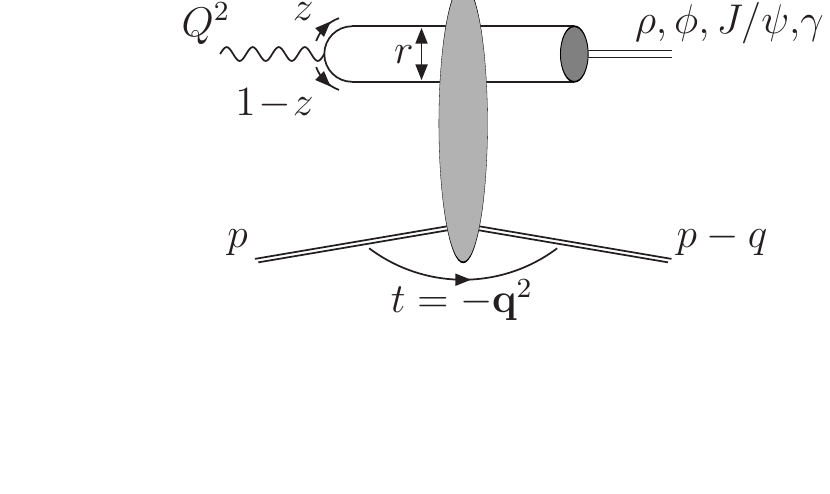}
\hfill
\includegraphics[width=0.45\textwidth]{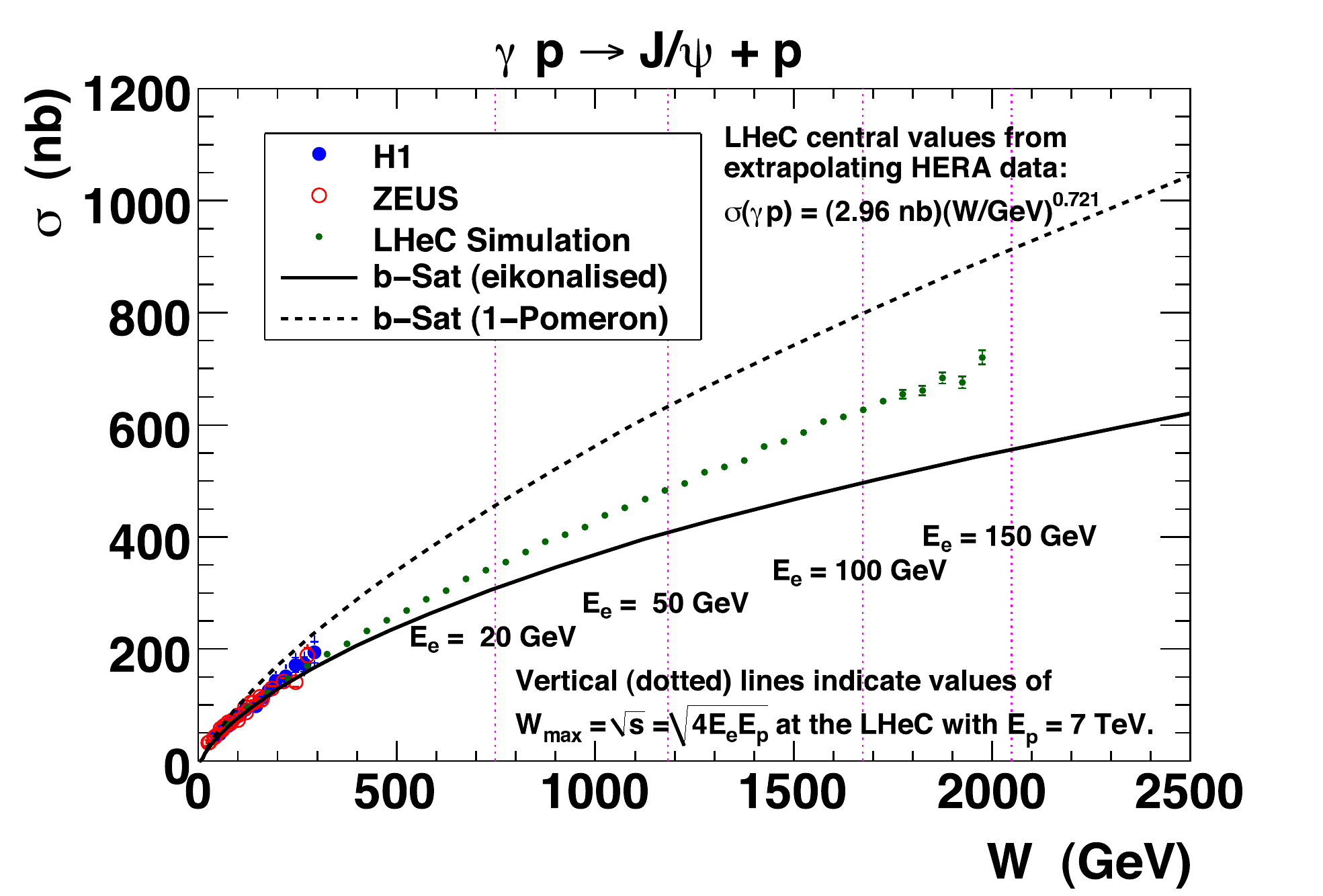}
\end{center}
\caption{Left: diffractive vector meson production in $\gamma^*\!+\!A$ scattering. Right: predictions (from \cite{AbelleiraFernandez:2012cc}) for the energy dependence of the total cross section at the LHeC, including or not non-linear effects.}
\label{dvmptotal}
\end{figure}

Diffractive vector meson production (or deeply virtual Compton scattering when the produced vector particle is a photon) is pictured in Fig.~\ref{dvmptotal} (left). This exclusive process provides access to the spatial distributions (and correlations) of partons in the transverse plane, unlike the more inclusive processes discussed before. Indeed, this information can be obtained through a Fourier transformation of the $t$ dependence of the cross section, where $t$ is the square of the momentum transferred by the target. Again, let me emphasize that there is no such direct access to spatial distribution in p+A collisions.

\begin{figure}[t]
\begin{center}
\includegraphics[width=0.9\textwidth]{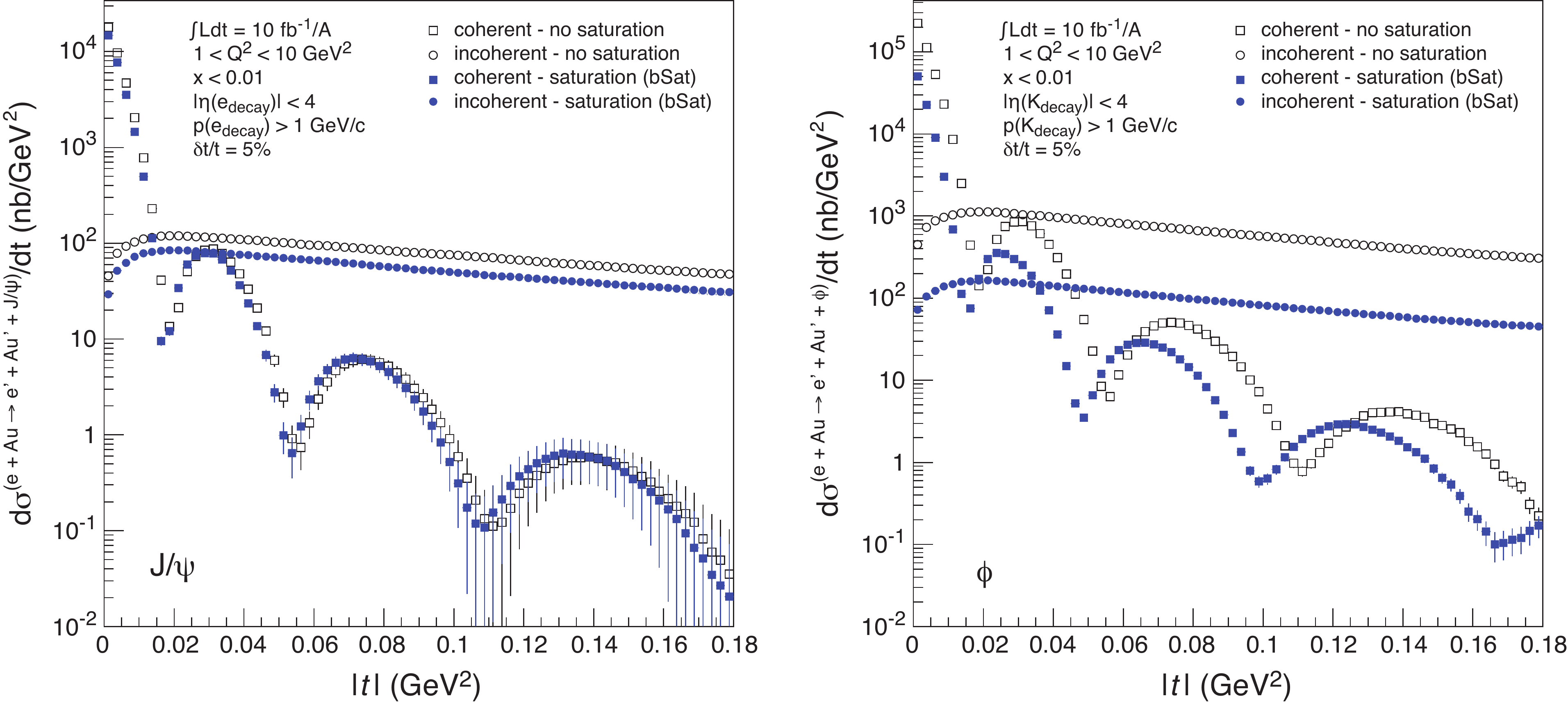}
\end{center}
\caption{Momentum transfer dependence of the cross section for diffractive $J/\Psi$ (left) and $\phi$ (right) production off a gold nucleus at the EIC. The coherent (intact gold nucleus, dominant at low $t$) and incoherent (dissociated gold nucleus, dominant at large $t$) productions are shown. The sensitivity to parton saturation is bigger in the case of the $\phi$ meson \cite{Toll:2012mb}.}
\label{dvmptdep}
\end{figure}

Fig.~\ref{dvmptotal} (right) shows, for a proton target, by how much the LHeC could extend the energy reach of HERA. Predictions for the energy dependence of the total cross section, including or not non-linear effects, are also displayed, showing that this measurement provides a clear smoking gun for non-linear effects already in the case of a proton target. Coming back to the momentum transfer dependence, Fig.~\ref{dvmptdep} shows diffractive $J/\Psi$ and $\phi$ production off a gold nucleus. Depending on how efficiently the incoherent background can be removed (incoherent diffraction allows the possible break-up of the target nucleus into its constituent nucleons), one could access the first minimum and maximum of the distribution.

\subsection{Di-hadron correlations}

The best experimental evidence of parton saturation so far has been observed at RHIC, looking at the azimuthal angle dependence of the correlation function of forward di-hadrons: the disappearance of the away-side peak in central d+Au collisions compared to p+p collisions \cite{Braidot:2010ig}. Qualitatively, the effect will be similar in e+A collisions, as explained below. However note that at the quantitative level, the di-hadron production process in e+A collisions involves a different operator definition of the unintegrated gluon distribution, and therefore is complementary to the RHIC measurement in p+A collisions \cite{Dominguez:2011wm}.

\begin{figure}[t]
\begin{center}
\includegraphics[width=0.95\textwidth]{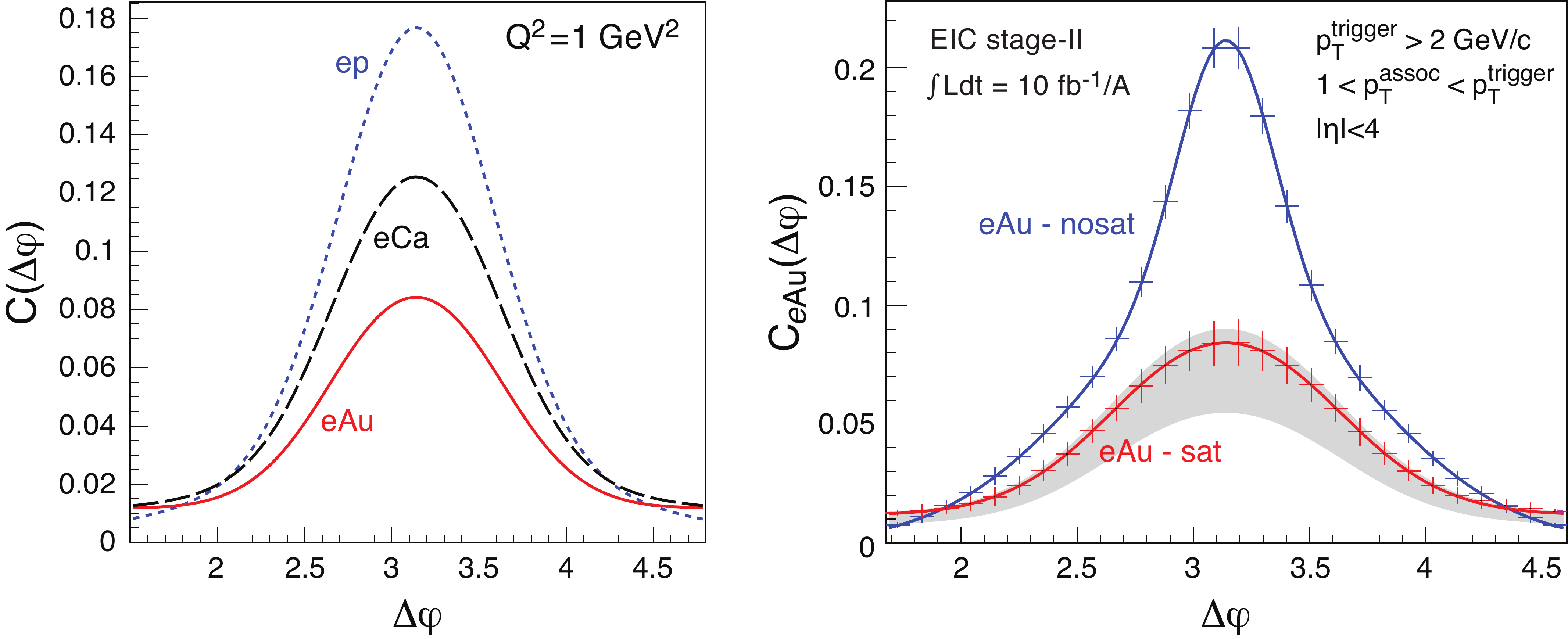}
\end{center}
\caption{Left: azimuthal angle dependence of the di-hadron correlation function in DIS off different nuclear targets. Right: illustration in the case of a gold nucleus, of the difference between the predictions with and without saturation effects, along with expected experimental errors at EIC energies (stage-2) \cite{Deshpande:2012bu}.}
\label{di-hadrons}
\end{figure}

In the e+A case, the two hadrons predominantly come from a quark and an antiquark, which were back-to-back while part of photon wave function. During the interaction, if they are put on shell by a single parton from the target carrying zero transverse momentum, as is the case when non-linear effects are not important, then the hadrons are emitted back-to-back (up to a possible transverse momentum broadening during the fragmentation process). By contrast, in the saturation regime, the quark and antiquark receive a coherent transverse momentum kick whose magnitude is of order $Q_s$, which depletes the correlation function around $\Delta\phi=\pi$, for hadron momenta not much higher than $Q_s$ \cite{Marquet:2007vb,Stasto:2011ru,Lappi:2012nh,yuan}. The effect is displayed in Fig.~\ref{di-hadrons} for EIC kinematics. It is striking in e+A collisions, more so than at RHIC due to the absence of a $\Delta\phi$-independent background, this provides yet another smoking gun for parton saturation.

\section{Conclusion}

The studies presented in this proceedings result from long-term work started years ago. They are a small sample of what has been accomplished to prepare a compelling physics case for future EICs. All detailed analysis can be found in (i) the e+A chapter of the INT report on the Physics case for the EIC \cite{Boer:2011fh}, edited by A. Accardi, M. Lamont and myself, (ii) the e+A chapter of the EIC white paper \cite{Deshpande:2012bu}, edited by Y. Kovchegov and T. Ullrich, and (iii) the small-x chapter of LHeC Conceptual Design Report \cite{AbelleiraFernandez:2012cc}, edited by N. Armesto, B. Cole, P. Newman and A. Stasto.

\section*{Acknowledgments}

This presentation would not have been possible without the help and hard work of the EIC task forces at Brookhaven and Jefferson lab, and of the LHeC small-x working group. This work is supported by the European Research Council grant HotLHC ERC-2011-StG-279579.


\end{document}